\documentstyle[12pt,preprint,aps]{revtex}
 
\begin{document}
\input{epsfig.sty}
 
\preprint{UCHEP-27\\ 
UMHEP-443}
\draft
\title{$\eta'$ production in $B$-decays: Standard Model vs. New Physics}
\author{Alexander L. Kagan}
\address{Department of Physics, 
University of Cincinnati \\
Cincinnati, OH 45221}
\author{
Alexey A. Petrov\footnote{
Address after September 1, 1997: Department of Physics and Astronomy,
The Johns Hopkins University, 3400 North Charles Str., 
Baltimore, MD 21218}}
\address{
Department of Physics and Astronomy \\
University of Massachusetts \\
Amherst MA 01003}
\maketitle
\begin{abstract}
\noindent

Standard model factorization can account for the large 
rate for $B^\pm \to \eta^\prime K^\pm$.  The large 
rate for $B \to \eta^\prime X_s$ is more problematic and requires a ``cocktail'' solution.
Individual contributions from factorization, $b \to \eta' s g$, and ``intrinsic charm''
fall significantly short of the observed rate.  In contrast, 
the observed $\eta^\prime$ yields are easily explained in models 
with ${\cal B}(b \to sg) \sim 15\%$. Factorization can 
account for the entire semi-inclusive $\eta^\prime$ yield without violating
rare decay constraints. Implications for $CP$ violation are discussed.  

\end{abstract}
\pacs{}

\section{Introduction}

The CLEO collaboration \cite{CLEOhawaii} has reported sizable 
production of fast $\eta '$ mesons,
\begin{equation} \label{cleoinc}
{\cal{B}}(B \to \eta'  X_s ) =  (7.5 \pm 1.5 \pm 1.1) \times 10^{-4} 
~(2.0 < p_{\eta'} < 2.7~GeV) 
\end{equation}
and a correspondingly large exclusive rate,
\begin{equation} \label{cleoexc}
{\cal B}(B^{\pm} \to \eta'  K^{\pm}) = (7.8 ^{+2.7}_{-2.2} \pm 1.0)
\times 10^{-5}.  
\end{equation}
There is no evidence for $\eta' K^*$ modes in the inclusive analysis, with most
events lying at large recoil mass, $m_{X_s}$. The experimental cut on 
$p_{\eta'}$ 
corresponds to $m_{X_s} <2.35~GeV$. 
There is also an exclusive branching fraction upper limit for
$\eta$ production,
\begin{equation} {\cal B}(B \to \eta K^\pm ) < .8 \times 10^{-5} 
(90\%~ c.l.).  \end{equation}

In the standard model (SM) factorization estimates for semi-inclusive 
\cite{deshheold,deshhe,ciuchini,browderetal} and 
exclusive \cite{palmer,deandrea} rare $B$ decay branching ratios ($Br$'s) are typically of 
order $10^{-4}$ and $10^{-5}$, respectively, 
so it is tempting to speculate 
that the large $\eta'$ yields are due to the intervention of new flavor physics.
A natural candidate to consider is 
enhanced $ b \to sg $ chromomagnetic dipole
operators which can explain several apparent discrepancies
between SM expectations and experimental results in $B$ decays:
the low semileptonic $Br$ and charm deficit
\cite{hou,implications,kaoncharm,ciuchini,nierste}, the kaon deficit \cite{kaoncharm}, 
and the low $\tau(\Lambda_b)/\tau(B_d)$ lifetime ratio
 \cite{neubert}.  All can be accounted for if 
${\cal B}(b \to sg) \sim 10\% - 15\% $, which is an interesting range 
because it is
naturally associated with $TeV$ scale dynamics for quark mass 
generation \cite{implications}.

In what follows we critically examine possible mechanisms for 
explaining the large $\eta '$ yields in the standard model and in models with enhanced
$b \to sg$.
We find that standard model factorization can, in principle, account for
the exclusive $\eta' $ yield.  For a liberal range of
factorization model parameters 
${\cal B}(B^\pm \to \eta' K^\pm)$ lies in the range
$(1.1 - 5.8) \times 10^{-5} $ for $m_s = .2~GeV$ and 
$(2.3 - 12.1)\times 10^{-5}$ for
$m_s = .1~GeV$.  It is worth noting in this regard 
that the recent values of $m_s$ obtained in lattice and QCD sum rule 
studies are quite low, e.g., $\overline{m_s} (2~GeV) \approx 128\pm 18~MeV$ \cite{msvalues1} and
$\overline{m_s} (2~GeV) \approx 100\pm 21 \pm 10~MeV$ \cite{msvalues2}. 
The inclusive yield is more problematic.
For the same range of parameters SM factorization\footnote{Throughout
we obtain results for inclusive $\eta'$ production for charged $B$ decays.
We will update our results in the near future to include the average over
charged and neutral $B$ decays.  
The branching ratio results are not significantly modified and our conclusions remain the same.}   
gives ${\cal B}(B^\pm \to \eta' X_s  ) 
\sim (0.5 - 2.5) \times 10^{-4}$ including the experimental cut, 
with the largest yields corresponding to a fairly limited region 
of parameter space. Of course there could be additional 
non-factorizable corrections, e.g., at the $10\% - 30\%$
level. 

Atwood and Soni (AS) \cite{atwood} have suggested
that the large inclusive rate is 
connected to the standard model QCD
penguins via the gluon anomaly, leading to the 
subprocess $b \to s g^* \to \eta' s g $. 
Taking a constant $g g \eta'$ vertex form factor
$H(0,0,m^2_{\eta'})$, extracted from $J/\Psi \to \eta' \gamma$,
they obtain agreement with Eq. (\ref{cleoinc}). 
More recently, Hou and Tseng (HT) \cite{hou2} have argued that the 
factor $\alpha_s$
implicit in $H$ should be running, which would lower 
AS's result by roughly a factor of 3.
However, both AS and HT have overlooked the 
leading $m_{\eta'}^2 /(q^2 - m_{\eta'}^2 ) $ dependence of the form factor, 
where $q$ is the virtual
gluon's momentum.  Including this dependence nominally reduces
AS's result to ${\cal B}(B \to \eta' X_s ) \sim 1.6 
\times 10^{-5} $ including the experimental cut, more than an order of magnitude 
below what is observed.

AS have estimated the contribution from 
decay of intermediate charmonia and from $\eta' - \eta_c$ mixing 
to be ${\cal B}(B \to \eta' X_s ) 
\sim 1.1  \times 10^{-4}$, including the cut.
Halperin and Zhitnitsky \cite{zhitnitsky}
have suggested that
the elevated glue content of the $\eta'$ gives it a large intrinsic charm 
component which simultaneously accounts for the 
exclusive and inclusive $\eta'$ yields via $b \to c \bar c s \to \eta' X $.
However, as already noted in \cite{hou2}, their prediction 
of ${\cal B}(B \to \eta' K ) \sim 2 
{\cal B}(B \to \eta' K^* )$ is inconsistent with the absence of $\eta' K^*$
modes in the inclusive analysis.  This alone implies
that the intrinsic charm $\eta'$ yields can not account for more than about a tenth of
Eqs. (\ref{cleoinc}) and 
(\ref{cleoexc}).\footnote{It was recently suggested 
\cite{chao} that the anomalously large $Br$ in Eq. (\ref{cleoinc})
can be explained by color-octet $c \bar c$ contributions
to $\eta'$ production, in analogy with the color-octet
mechanism in the NRQCD description of charmonium production. 
However, the $c \bar c$ pair must be far off shell 
inside of the $\eta'$, invalidating the use of the NRQCD formalism.}  
Finally, as a by-product of their SM analysis of $b \to s g g$ 
Wyler and Simma have obtained an ${\cal O}(10^{-6})$ $Br$ estimate for 
$B \to K + glueball$ \cite{wyler}, 
indicating that the $Br$ for $b \to s gg \to s \eta' $ should not exceed 
${\cal O}(10^{-5})$.
In our view, although each of the above mechanisms falls short by itself, 
a standard model `cocktail solution' for the inclusive $\eta' $ yield 
is still possible given the large hadronic uncertainties involved,
with factorization providing the largest contribution. 

In contrast, we find that models with enhanced $b \to sg$ readily
account for the observed inclusive yield in the factorization model alone,
obtaining ${\cal B}(B \to \eta' X_s) \approx (1.0 - 9.0) \times 10^{-4}$.
The actual value largely depends on the choice of weak phases in the dipole 
operator coefficients,
and $B \to \eta'$ form factors.
Moreover, we have checked that the large $\eta'$ yields 
are compatible with
the potentially restrictive CLEO
upper limits on $B \to \phi X_s $ \cite{CLEOphiX} and $B \to \phi K$ \cite{CLEOphiK} 
for a significant range of parameters in the factorization approximation. 
The recoil spectrum increases with $m_{X_s}$, in qualitative agreement 
with observation.
In the case of the anomaly mediated process $b \to \eta' sg $, 
HT argued that the entire inclusive yield can be reproduced 
if $b \to sg$ is enhanced.  Taking into account the previously 
overlooked $m_{\eta'}^2 /(q^2 - m_{\eta'}^2 )$ dependence of the $g g \eta'$ 
form factor, but ignoring the subleading $q^2$ dependence of 
$\alpha_s$ in HT, we find 
${\cal B}(B \to \eta' X_s ) \sim (0.3 - 1.6) \times 10^{-4} $
for ${\cal B}(b \to sg) \approx 15\%$.  Although significantly 
smaller than the rates obtained in the factorization approach
it is clear that the sum of the two
can easily reproduce Eq. (\ref{cleoinc}) if $b \to sg$ is enhanced.

An important feature of quark model $B$ decays 
pointed out by Lipkin 
\cite{lipkin} is the
interference between spectator ($b \to s \bar u u $)
and non-spectator ($b \to s \bar s s$) production of $\eta'$ and 
$\eta $ mesons.
It is constructive for exclusive $\eta' K $ modes and destructive for 
exclusive $\eta K$ modes, leading
to order of magnitude smaller rates for the latter, in accord 
with the recent measurements. We have checked this explicitly in the 
factorization approximation for the standard model and for 
enhanced $b \to sg$. 
The constructive interference can also account for the larger
rates for $\eta' K$ versus $K \pi , \phi K$.
Lipkin has noted that the same interference pattern persists generally for
orbital parity even final states, e.g., constructive
for $\eta' K^{**}$
and destructive for $\eta K^{**}$, while the reverse is true 
for orbital parity odd final states, e.g., destructive for $ \eta' K^* $ and
constructive for $\eta K^* $. 
This most likely can explain the absence of 
$ \eta' K^{* \pm} $ events in the inclusive analysis. 
We leave a factorization model analysis 
of $B \to \eta' K^*$ for future work.

The weak phases associated with new physics contributions to the 
dipole operator coefficients 
generally lead to
large $CP$ violating asymmetries in charmless $B$ decays in
the factorization approximation \cite{kaganhawaii}
when strong phases in the penguin amplitudes from $c\bar{c}$ rescattering 
\cite{bandersilvsoni,hougerard} are taken into account. 
For SM factorization we obtain CP asymmetries in the range 
$4.5\% - 8\%$ for $B^\pm \to \eta' K^\pm$ and $2\% - 5 \%$ for 
$B \to \eta' X_s $ with, as
usual, the larger asymmetries corresponding to smaller $Br$'s.
Including enhanced $b \to sg$ can give asymmetries of order
10\% (exclusive) and 5\% (inclusive) for $Br$'s in the observed range.
These are not dramatically larger than the SM asymmetries, 
unlike the `pure penguin' processes $B^{\pm} \to \phi K^{\pm}$ and 
$B \to K^0 \pi^{\pm}$ for which large asymmetries are possible
compared to $\leq 1\%$ in the standard model. 
We note that enhanced $b \to sg$ can also lead to large 
isospin violating asymmetries in radiative $B$ decays \cite{alexey}.

In the next section we discuss $\eta'$ production in the standard 
factorization approach.  In Section 3 we focus on the gluon anomaly subprocess
and comment on other `exotic' mechanisms.
We conclude with discussion of our results in Section 4.

\section{$\eta'$ and $\eta$ Production in the Factorization Approach.}

The effective Hamiltonian for non-leptonic charmless $b \to s$ transitions
is given below\footnote{The sign of the operators $O_{11}$, $O_{11}'$
corresponds to use of $- i g_s T^a $ for the quark - gluon vertex Feynman rule 
\cite{burasfleischer}, leading to destructive interference in the standard model
between penguin and chromomagnetic dipole operator contributions to charmless 
$b \to s q \bar q$ decays.}

\begin{eqnarray}
{\cal H}_{eff} = \frac{4 G_F}{\sqrt{2}} 
\Bigl [ V_{ub} V_{us}^* \sum_{i=1}^{2}
c_i O_i^u &-& V_{tb} V_{ts}^* (\sum_{i=3}^{11} c_i O_i + c_{11}' O_{11}' )
\Bigr ]~~~, \nonumber \\
O_1^u = \bar s_\alpha \gamma_\mu L u_\beta \bar u_\beta 
\gamma^\mu L b_\alpha~&,&~
O_2^u = \bar s \gamma_\mu L u
\bar u \gamma^\mu L b~,~ \nonumber \\
O_{3(5)} = \bar s \gamma_\mu L b
\sum_q \bar q \gamma^\mu L (R) q~&,&~
O_{4(6)} = \bar s_\alpha \gamma_\mu L b_\beta 
\sum_q \bar q_\beta \gamma^\mu L (R) q_\alpha~,~ \nonumber \\ 
O_{7(9)} = \frac{3}{2} \bar s \gamma_\mu L b \sum_q
e_q \bar q \gamma^\mu L (R) q~&,&~
O_{8(10)} = \frac{3}{2} \bar s_\alpha \gamma_\mu L b_\beta 
\sum_q e_q \bar q_\beta \gamma^\mu L (R) q_\alpha~, 
\nonumber \\ 
O_{11} = \frac{g_s}{32 \pi^2} m_b (\mu)
\bar s \sigma_{\mu \nu} R t^a b G^{\mu \nu}_a~&,&~ 
O_{11}' = \frac{g_s}{32 \pi^2}
m_b (\mu) \bar s \sigma_{\mu \nu} R t^a b G^{\mu \nu}_a~, 
\end{eqnarray}
where the standard notations are used and $q$ is summed over $u,d,s$.

In our numerical analysis we take 
$\alpha_s (M_Z) = .117 $, $m_t = 174~GeV$, $m_b = 5.0~GeV$, $m_c = 1.63~GeV$ 
(HQET parameter $\lambda_1 = -.2 $), and $m_s = .1$ to $.2~GeV$. 
For KM matrix entries we take $V_{ts} V_{td}^* = .037$,
and Wolfenstein parameters $\rho = -.15$, $\eta = .33$ which are 
in the range favored by the SM analysis of Ali and London \cite{london}.
We use the Next-to-Leading Order (NLO) scheme-independent 
effective Wilson coefficients 
\cite{fleischer,deshheold,palmer} at $\mu = m_b$, obtaining
\begin{eqnarray} 
c_1  = - 0.306,~~c_2 = 1.15,~~c_3 = 0.017 - 0.37 P_s,~~
c_4 = - 0.037 + 1.11  P_s,
\nonumber \\
c_5 = 0.01 - 0.37 P_s,~~
c_6 = -0.045 + 1.11 P_s,~~
c_7 = 1.03 \times 10^{-5} + 0.33 P_e,
\nonumber \\
c_8 = 3.82 \times 10^{-4},~~
c_9 = -0.01 + 0.33 P_e,~~c_{10} = 1.97 \times 10^{-3}. 
\end{eqnarray}
The $P_{s,e}$ are given by $P_s = {\alpha_s \over {8 \pi}}({ 10 \over 9 } - 
G(m_c,\mu,q^2))$, $P_e = {\alpha_{em} \over{8 \pi}} ({ 10 \over 9 } - 
G(m_c,\mu,q^2))$,
where 
\begin{equation} 
G(m,\mu,q^2) = - 4 \int_0^1 x (1 -x) 
\ln {{m^2 - x(1-x) q^2}\over \mu^2} dx . 
\end{equation}
$q$ is the momentum of the virtual gluon in the penguin diagram.  $G$ becomes imaginary when 
$q^2 >  4 m_c^2 $ and a strong phase 
is generated \cite{bandersilvsoni,hougerard}.

The Leading Order (LO) dipole coefficient is $c_{11}^{sm}(m_b) 
\approx  -.295$, which corresponds
to ${\cal B}(b \to sg) \sim .2 \% $. 
For enhanced dipoles we'll consider ${\cal B}(b \to sg) \approx 15\% $, 
corresponding 
to\footnote{This value is determined using the total NLO standard model,  
inclusive $b $ decay width, see next footnote, and
the LO expression for ${\cal B}(b \to sg)$.} 
 $(|c_{11}|^2 + |c_{11}'|^2 )^{1 \over 2} \approx 2.26$. In obtaining $Br$'s we 
consider two total $B$ decay widths: the NLO SM value\footnote{This 
includes the NLO corrected $b \to c$ and $b \to u$ transitions \cite{bagan}, 
taking into account ${\cal O}(\alpha_s c_2^2)$ 
corrections to the LO penguin contributions to
$b \to c \bar c s $, and ${\cal O}(1/m_b^2)$ HQET corrections \cite{bigietal}. 
The charmless $b \to s$ transitions and $O_{11}$ have not been included
but their
total correction is $\sim 1\%$. } for the above inputs,
$\Gamma_{SM} \approx 3.3 \times 10^{-13}~GeV$, and 
$3.88 \times 10^{-13}~GeV$ for ${\cal B}(b \to sg) \approx 15\%$. 
The dipole coefficients are parametrized as
\begin{equation} c_{11} = -|c_{11}| 
e^{i \theta_{11}},~~~c_{11}' = -|c_{11}'| e^{i \theta_{11}'}.
\end{equation}
Generally, large direct $CP$ violation effects result
from interference between new amplitudes containing 
the weak phases $\theta_{11},~\theta_{11}'$
and standard penguin amplitudes containing 
strong phases from the $G$ function.

In what follows the dipole operators are included by allowing the 
off-shell gluon to turn into a quark-antiquark pair. 
For $O_{11}$ this leads to the effective Hamiltonian
\begin{equation}
{\cal H}_{11} = - \sum_{q=u,d,s} i \frac{G_F}{\sqrt{2}} 
\frac{\alpha_s}{\pi q^2}
V_{tb} V_{ts}^* C_{11} m_b (\mu) \bar s(p_s) \sigma_{\mu \nu} R t^a
b(p_b) \bar q(p_2) \gamma^\mu t_a q(p_1) q^\nu,
\end{equation}
where $q^2 \equiv (p_b - p_s)^2 $.  In the standard model $H_{11}$ interferes
destructively with the penguin amplitudes.
In practice, we take $\alpha_s = g_s (m_b) g_s (q)/4 \pi $,
identifying one factor of $g_s $ with the virtuality of the gluon.  This
procedure models the fact that in approaching the `on-shell'  
gluon limit  ${\cal H}_{11}$ 
becomes increasingly non-local so that gluon fragmentation must 
take place at non-perturbative size scales.
Performing Fiertz rearrangement yields\footnote{This result differs
from the result in \cite{deshhe}.} 
\begin{eqnarray}
{\cal H}_{11} = + \sum_{q=u,d,s} \frac{G_F}{\sqrt{2}} 
\frac{\alpha_s (q^2)}{4 \pi q^2}
V_{tb} V_{ts}^* C_{11}^{(\prime)} m_b (\mu)
\frac{N_c^2-1}{N_c^2} \Biggl [ \delta_{\alpha \beta} 
\delta_{\alpha' \beta'} -
\frac{2 N_c}{N_c^2-1} t^a_{\alpha \beta} t^a_{\alpha' \beta'} 
\Biggr ] \nonumber \\ 
\Bigl \{2 m_b \bar s_\alpha \gamma_\mu L q_\beta \bar q_{\alpha'} 
\gamma^\mu L b_{\beta'}
- 4 m_b \bar s_\alpha R q_\beta
\bar q_{\alpha'} L b_{\beta'} +
2 m_s \bar s_\alpha \gamma_\mu R q_\beta \bar q_{\alpha'} 
\gamma^\mu R b_{\beta'}
- \nonumber \\ 4 m_s \bar s_\alpha L q_\beta
\bar q_{\alpha'} R b_{\beta'} +
(p_b + p_s)_\mu \Bigl [
\bar s_\alpha \gamma_\mu L q_\beta
\bar q_{\alpha'} R b_{\beta'} +
\bar s_\alpha R q_\beta
\bar q_{\alpha'} \gamma^\mu R b_{\beta'} + \nonumber \\ i \bar s_\alpha
\sigma^{\mu \nu} R q_\beta \bar q_{\alpha'} \gamma_\nu R b_{\beta'} - i \bar
s_\alpha \gamma_\nu L q_\beta
\bar q_{\alpha'} \sigma^{\mu \nu} R b_{\beta'} \Bigr ] \Bigr \}
\end{eqnarray}
The effective Hamiltonian 
${\cal H}_{11}^\prime$ corresponding to $O_{11}'$
is obtained by substituting $C_{11} \to C_{11}'$ and $L \to R$ in the above.
In all of our factorization model results we take $N_c = 3$.

\subsection{Semi-inclusive $\eta',~\eta$ production.}

Factorization model contributions to $B^- \to \eta^{(\prime)} X_s$ 
involve two classes of amplitudes distinguished by their
hadronization pattern, which we loosely refer to as ``two-body'' 
and ``three-body'' decays.\footnote{A similar treatment of $B \to KX$
is presented in \cite{browderetal}.} The matrix element is of the form
\begin{equation}  M = M^{2b} + M^{3b} \end{equation} 
so that
\begin{equation} |M|^2 = |M^{2b} |^2 + |M^{3b}|^2 + 
M^{2b \dagger} M^{3b} +  M^{3b \dagger} M^{2b}.
\end{equation}
In the two-body decays an $\eta'$ or $\eta$ is formed from an $s \bar s $ 
pair via the subprocess $b \to s \bar s s$,  
which in the parton model looks like
$b \to \eta^{(\prime)} s $. Applying the Gordon identity and Dirac 
equation the matrix element for $\eta^{(\prime)}$ production can be 
written in the simplified form
\begin{equation}
M^{2b}  =
\zeta^{\eta^{(\prime)}}_1  \langle X_s | \bar s L b | \overline{B} \rangle + 
\zeta^{\eta^{(\prime)}}_2 
\langle X_s | \bar s R b | \overline{B}\rangle. 
\label{eq:2body} 
\end{equation}
For $\eta'$ production the coefficients depend on
the matrix elements 
\begin{mathletters}
\label{etapr:all}
\begin{equation} 
\langle \eta' |\bar s \gamma^\mu \gamma_5 s |0 \rangle =
- i p^{\mu}_{\eta'} {\sqrt{6}\over 3} (F_0 \cos \theta - 
\sqrt{2} F_8 \sin \theta )
\label{etapr:a}
\end{equation}
\begin{equation}
\langle \eta' |\bar s  \gamma_5 s |0 \rangle  =  i \sqrt{3 \over 2}
{m_{\eta'}^2 \over 2 m_s} F_8 \sin \theta 
\label{etapr:b}
\end{equation}
\begin{equation}
\langle \eta' |\bar u \gamma^\mu \gamma_5 u |0 \rangle 
= - i p^{\mu}_{\eta'} \sqrt{{2 \over 3}} (F_0 \cos \theta + 
{1 \over \sqrt{2}} F_8 \sin \theta ),
\label{etapr:c}
\end{equation}  
\end{mathletters}
where $F_0$ and
$F_8$ are the decay constants
for the $SU(3)_F$ singlet and octet axial vector currents,
and the anomaly has been taken into account in 
$\langle \eta' |\bar s  \gamma_5 s |0 \rangle$ 
following \cite{tytgat}.
For the $\eta$ the corresponding matrix elements are  
\begin{mathletters}
\label{eta:all}
\begin{equation}
\langle \eta |\bar s \gamma^\mu \gamma_5 s |0 \rangle =
 i p^{\mu}_{\eta} {\sqrt{6}\over 3} (F_0 \sin \theta + 
\sqrt{2} F_8 \cos \theta )
\label{eta:a}
\end{equation}
\begin{equation}
\langle \eta |\bar s  \gamma_5 s |0 \rangle  =  i {\sqrt{3 }\over 2}
{m_{\eta}^2 \over  m_s} F_8 \cos \theta
\label{eta:b}
\end{equation}
\begin{equation}
\langle \eta |\bar u \gamma^\mu \gamma_5 u |0 \rangle 
=  i p^{\mu}_{\eta} \sqrt{{2 \over 3}} 
(F_0 \sin \theta - {1 \over \sqrt{2}} F_8 \cos \theta ).
\label{eta:c}
\end{equation}
\end{mathletters}
Phenomenological fits in \cite{tytgat} give 
$F_0 \approx F_8 \approx F_\pi $, where $F_\pi = 132/\sqrt{2} ~MeV$, and
an $\eta' - \eta$ mixing angle $\theta_{\eta \eta'} \approx - 17^o $.
Chiral perturbation theory favors $F_8 \approx 1.25 F_\pi $ and 
$\theta_{11 } \sim - 21^o $ \cite{CPT,tytgat}. $Br$'s are given for both choices. 
We make the usual quark model approximation, taking 
equal momenta for the two $s$ quarks in the $\eta^{(\prime)} $, which gives
\begin{equation} q^2 = m_b^2 /2  - m_{\eta^{(\prime)}}^2 /4 + m_s^2 /2 . 
\end{equation}
For simplicity
we take the $b$ quark and spectator at rest so that the $\eta^{(')}$ 
energy is fixed. 
Taking their momenta into account as in Ref. \cite{altarelli}
would not have a significant impact on the overall $Br$ but would 
smear the ``two-body'' $\eta^{(')}$ spectrum by a few hundred $MeV$.

SM $b \to \eta' s $ $Br$'s can be found in Table \ref{first}. Associated CP 
asymmetries are $\sim  + 1\% $.  The large $m_s$ dependence is 
traceable to the $1/m_s $ factor in 
$\langle \eta' |\bar s  \gamma_5 s |0 \rangle$ which, as we'll see shortly, 
can lead to large exclusive $Br$'s in the SM.  The corresponding 
$b \to \eta  s$ $Br$'s range from $(3.4- 9.0 ) \times 10^{-5}$, 
with CP asymmetries of $\sim - 1\%$.
Enhanced $b \to sg$ can increase these contributions by a factor 
of 2 to 3, but the dominant contribution will come from the 
``three-body'' decays.

The ``three-body'' decays involve hadronization of the spectator 
into the $\eta^{(\prime)}$ via the subprocess $b\to s \bar u u$.
In the parton model this is interpreted as 
$B^- \to \eta^{(\prime)} s \bar u $, with $p_X = p_s + p_{\bar u} $.
Using the equations of motion 
and again, for simplicity, taking the $b$ quark at 
rest\footnote{There are in general
two additional terms which vanish in this limit, being  proportional to 
$p_B^\mu p_b^\nu  \bar s 
\sigma_{\mu\nu} R u $ and
$p_B^\mu p_b^\nu  \bar s 
\sigma_{\mu\nu} L u $ following application 
of the equations of motion(EOM).}, the matrix element can be 
simplified to the form
\begin{eqnarray} 
M^{3b} = {\xi^{\eta^{(\prime)}}_1}^\mu   \langle X_{s \bar u} | 
\bar s \gamma_\mu L u | \overline{B} \rangle
+ \xi^{\eta^{(\prime)}}_2 \langle X_{s\bar u} | 
\bar s R u | \overline{B} \rangle \nonumber \\
+ {\xi^{\eta^{(\prime)}}_3}^\mu  
\langle X_{s \bar u} | \bar s \gamma_\mu R u | \overline{B} \rangle
+ \xi^{\eta^{(\prime)}}_4 \langle X_{s \bar u} | \bar s  L u | 
\overline{B} \rangle 
\label{eq:3body}
\end{eqnarray}
The matrix elements entering the $\xi_i$ are parameterized as 
\cite{BSW}:
\begin{mathletters}
\label{betap:all}
\begin{equation} 
\langle \eta^{(\prime)} | \bar u \gamma^\mu  b | B^- \rangle =
(p_B +p_{\eta^{(\prime)} })^\mu F_{1^-}^{\eta^{(\prime)}} 
(m_{X_s}^2) + (p_B - p_{\eta^{(\prime)}})^\mu
 f^{\eta^{(\prime)}} (m_{X_s}^2)
\label{betap:a}
\end{equation}
\begin{equation}
\langle \eta^{(\prime)} | \bar u   b | B^- \rangle = 
{1\over {m_b - m_u}} (m_{B^-}^2 - m_{\eta^{(\prime)}}^2 ) 
F_{1^-}^{\eta^{(\prime)}} (m_{X_s}^2)
+ m_{X_s}^2  f^{\eta^{(\prime)}} (m_{X_s}^2)
\label{betap:b}
\end{equation}
\begin{equation}
\langle \eta' | \bar u \sigma^{\mu\nu}  b | B^- \rangle = 
-i  {{(F_{1^-}^{\eta^{(\prime)}} (m_{X_s}^2) - 
f^{\eta^{(\prime)}} (m_{X_s}^2) )}\over {2 m_b}}
(p_B^\mu p_{\eta^{(\prime)}}^\nu - p_B^\nu p_{\eta^{(\prime)}}^\mu ),
\label{betap:c}
\end{equation}
\end{mathletters}
where,
\begin{equation}  f^{\eta^{(\prime)}} (p^2) = \frac{m_{\eta^{(\prime)}}^2 - 
m_{B^-}^2}{p^2} 
[F_{1^-}^{\eta^{(\prime)}} (p^2) - F_{0^+}^{\eta^{(\prime)}} (p^2) ].
\end{equation}
With a simple monopole parameterization assumed for the form factors
they are determined by fixing $F_{1^-}$ and $F_{0^+}$ at zero momentum
transfer, where they must be equal.   
It appears, based on various models in the literature 
\cite{BSW,deandrea,stechneubNRX,stechneub}, 
that a reasonably broad
range to consider for all $B$ to light pseudoscalar form factors
is $F_{1^-} (0) = .25 - .5$. 

\begin{figure}
\centerline{
\epsfbox{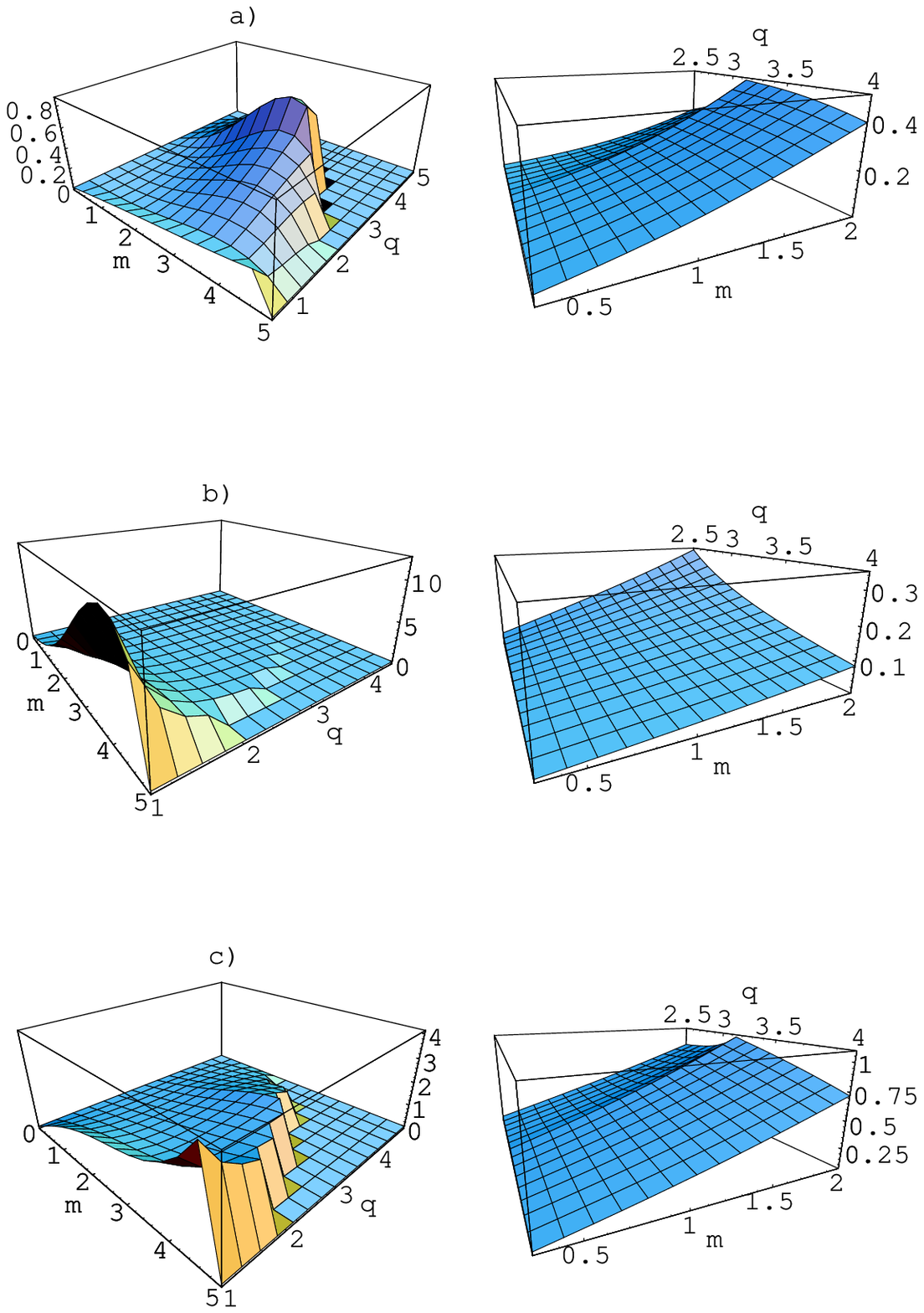}}
\caption {$10^3 { {d {\cal B} }\over {dm dq}  } (b \to s \bar d d) $   
for:  (a) SM,
(b) Enhanced $O_{11}$ by itself, (c) SM with enhanced 
$O_{11}$ for $\theta_{11} = 180^o$ (constructive interference).  
The regions relevant to two-body or quasi two-body decays in the 
factorization approximation are blown up.}
\label{one}
\end{figure}

To gain insight into the factorization
approach it is useful to study the quark level NLO Dalitz plots for  
$b \to s \bar q q$ shown in Fig. (\ref{one}), where  
$q^2 \equiv (p_b - p_s)^2 $ and $m^2 \equiv (p_s + p_{\bar q})^2$.  For simplicity we 
only consider $b \to s \bar d d$, which does not contain tree-level 
contributions.  The same conclusions can be drawn from
the Dalitz plots for $b \to s \bar u u$ and $b \to s \bar s s$.  
We have blown up the regions relevant to factorization 
for two-body or semi-inclusive ``quasi'' two-body decays, 
i.e., $q^2 \sim m_b^2 /2 $ and $m \leq 2~GeV$.  
It is clear why models with enhanced $b \to sg$ readily evade constraints 
from rare $B$ decays in the factorization model.  Although 
the charmless $Br$ is an order of magnitude larger than in the SM 
it is peaked at low $q^2$,
i.e., the ``on-shell'' gluon 
limit.\footnote{The singular low $q^2 $ dependence of $b \to s \bar q q$
originates from the pure dipole contributions, e.g., ${\cal O}(c_{11}^2)$. Although
these contributions come in at ${\cal O}(\alpha_s^2)$ they are
clearly important in the case of enhanced dipole coefficients and should be included,
particularly since we will be interested in 
comparison of inclusive quark level and factorization model Dalitz plots.
This dependence will be canceled 
in the total charmless $Br$
by ${\cal O}(\alpha_s )$ corrections to $b \to sg$.}
In the region relevant to rare decays, e.g., 
$B \to K\pi, K \phi, \phi X_s$, etc., 
the pure dipole amplitude is only of same order as in the standard model.
This means that in general large interference is
possible between the penguin and dipole amplitudes.  
Generally this will lead to large CP violation \cite{kaganhawaii}
because at $q^2 \sim m_b^2/2$ the penguin amplitude should develop 
large strong phases \cite{bandersilvsoni,hougerard}, and the dipole amplitude 
can have arbitrarily large weak phases.\footnote{ 
It is also clear that it is not possible to obtain an ${\cal O}(10\%)$ 
charmless $Br$ via new physics contributions to the penguin operators 
without violating 
rare $B$ decay constraints, since this would necessarily increase the
differential $Br$'s in the factorization region by an order of 
magnitude or more
relative to the SM.  For example, 
${\cal B}(B \to \phi X_s)$ would exceed the CLEO upper limit
by a factor of 20 - 40 in the factorization approach.}

\begin{figure}
\centerline{
\epsfbox{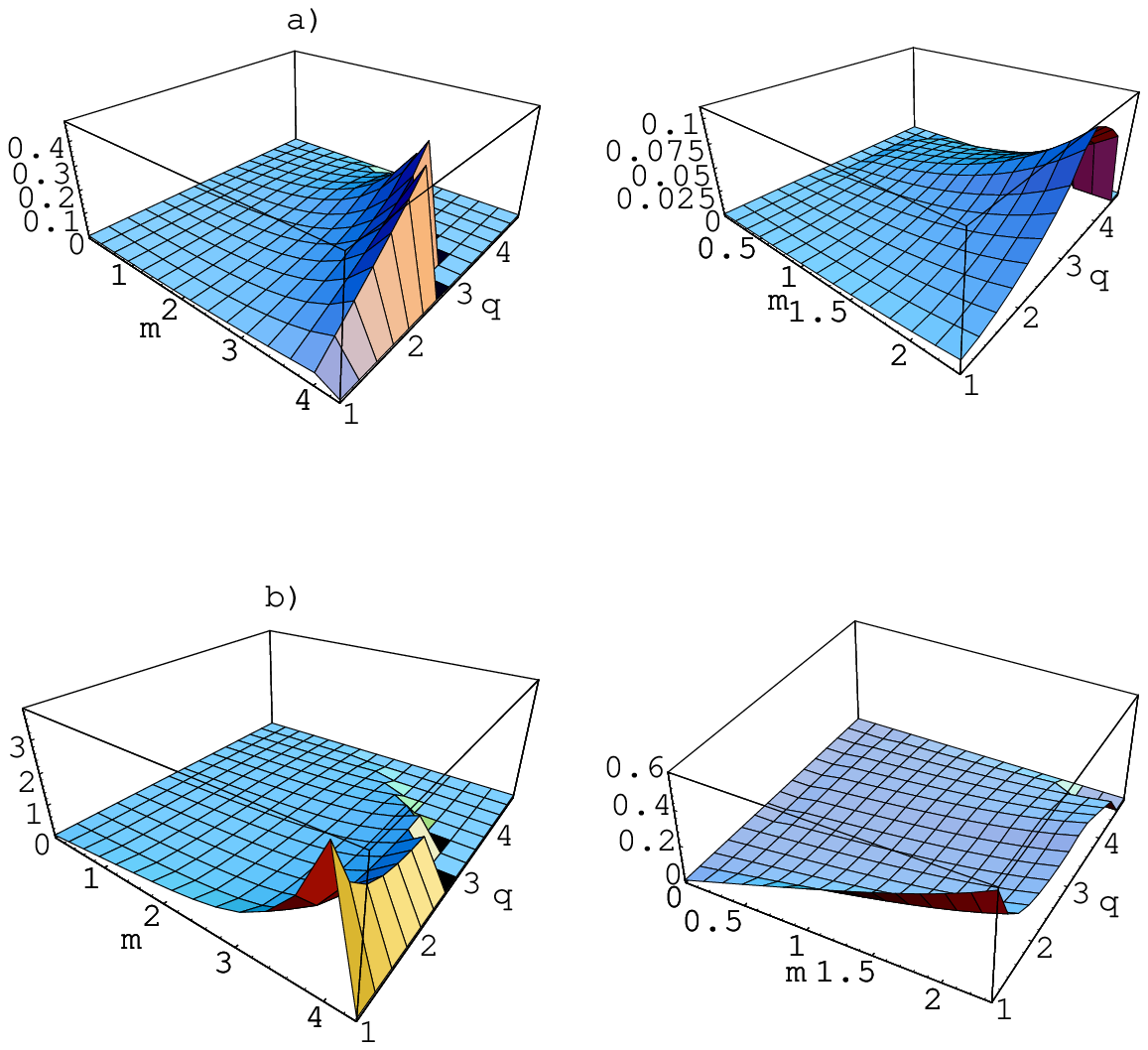}}
\caption {$10^3 { {d {\cal B} }\over {d m_{X_s} dq}  } 
(B^-  \to \eta'  s \bar u ) $   
for:  (a) SM, (b) SM with enhanced $O_{11}$ for $\theta_{11} = 180^o$ 
(constructive interference).  
The regions within the experimental cut are blown up.}
\label{two}
\end{figure}

\begin{figure}[t]
\centerline{
\epsfbox{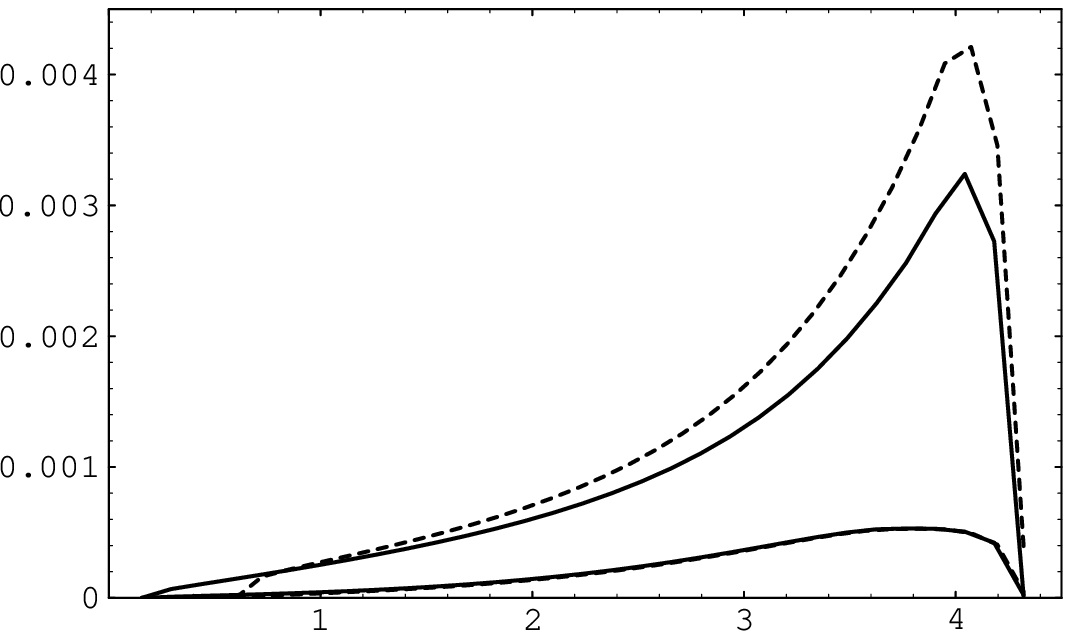}}
\caption {$10^3 { {d {\cal B} }\over {dm_{X_s}} } 
(B^-  \to \eta'  s \bar u ) $ for SM (lower curves), and SM with 
enhanced $O_{11}$ and $\theta_{11} = 180^o$ (upper curves).  
Solid curves are for $m_s = .15~GeV$, $m_u = .005~GeV$,
dashed curves are for $m_s = .5~GeV$ and $m_u = .1 ~GeV$.}
\label{three}
\end{figure}

Three-body $B^- \to \eta' s \bar u$ Dalitz plots
are shown in Fig. (\ref{two}).  Note that the `quark level' 
definition for $q$ is used, i.e., $p_b - p_s$ rather than $p_B - p_s$.  
The plots do loosely resemble the corresponding inclusive
plots in Fig. (\ref{one}).  The regions surviving the cut 
$m_{X_{s}} < 2.35~GeV$ are blown up.
The corresponding recoil spectra, shown in Fig. (\ref{three}), 
show a characteristic rise 
with $m_{X_{s }}$, like the measured spectrum \cite{CLEOhawaii}.
Including the two-body spectra and interference between two-body  
and  three-body $\eta'$ production  (2b-3b interference), particularly
the destructive interference expected for exclusive $\eta' K^* $ 
production \cite{lipkin}, should improve agreement with 
the shape of the observed spectrum at low $m_{X_s}$.

\begin{figure}
\centerline{
\epsfbox{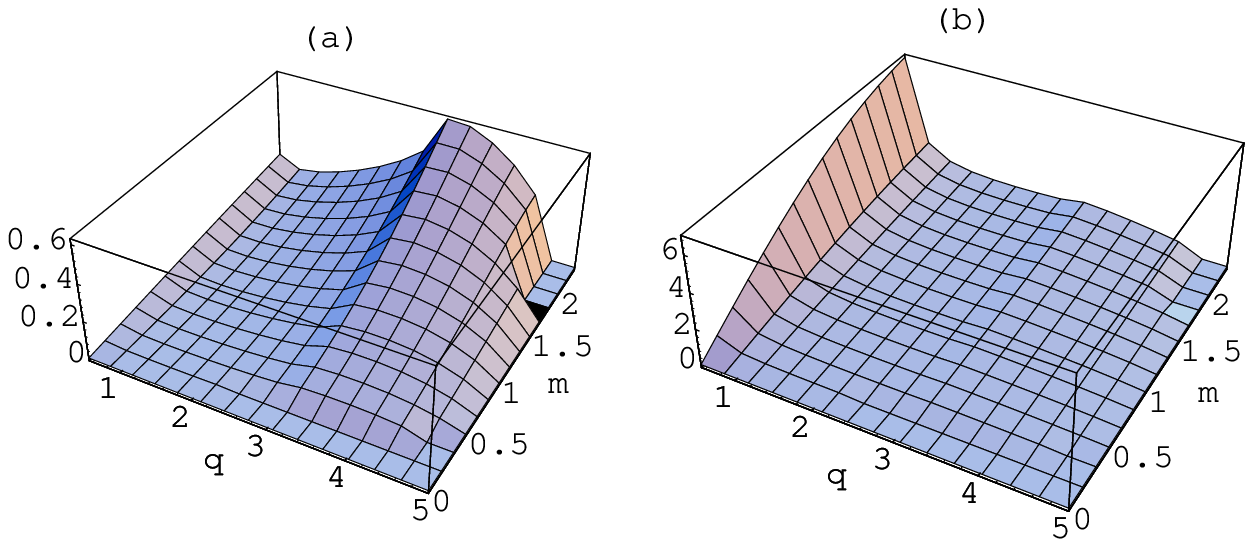}}
\caption {$10^3 { {d {\cal B} }\over {d m dq}} (b   \to s \bar d d )$   
for:  (a) SM, (b) SM with enhanced $O_{11}$ and 
$\theta_{11} = 180^o$, in the regions relevant to $\eta'$ production 
with experimental cut.}
\label{four}
\end{figure}

To address the question of whether 2b-3b
interference can have a significant 
impact on the {{\it semi-inclusive}} $\eta^{(\prime)}$ yields we 
have blown up the regions of Figs. (\ref{one}a) and 
({\ref{one}c}) which are relevant for three-body $\eta'$ 
production, with experimental cut, in Fig. (\ref{four}). The three-body amplitude can 
interfere with two-body $\eta^{(\prime)}$ production when
$q \sim m_{\eta^{(\prime)}}$ and $m_{X_s} \sim 1~GeV$.  
A glance at Fig. (\ref{four}a) will convince the reader that
in the SM the relative contribution of this region to the 
overall three-body $Br$ is not important, 
so that 2b-3b interference should have little impact on the 
relative $\eta'$ vs. $\eta$ yields.
However, hierarchies of 2 to 4 between the $\eta'$ and $\eta$ yields may
result from reasonable differences between the $B \to \eta$ and 
$B \to \eta'$ form factors. In the case of enhanced $b \to sg$,
Fig. (\ref{four}b), the relative contribution
of this region to the three-body 
yield appears to be significant, particularly for $\eta$ production, 
due to the singular 
behaviour at low $q^2$, so that the interference could turn out to be relevant.
For now only the pure two-body and three-body factorization 
estimates will be considered, with a low $q^2 $ cutoff imposed on the latter. 
We make no pretense that factorization is valid at low $q^2$ for dipole 
operator contributions, recognizing that this 
further contributes to the uncertainty
in $\eta'$ production in models with large $b \to sg$. 

In Table \ref{second} we study the dependence 
of ${\cal B}(B^- \to \eta' s \bar u)$
with experimental cut on choice of low $q^2 $ cutoff.  The singular 
behavior is apparent below $1~GeV^2$ for enhanced $b \to sg$, 
while there is relatively little change 
in the SM $Br$'s below $1~GeV^2$.  The three-body $B \to \eta$ $Br$'s are 
very close to the $\eta'$ yields in the SM 
and $\sim 5\% - 15\% $ larger with enhanced $b \to sg$.
In the ensuing discussion we fix the $q^2$ cutoff at $1~GeV^2$ 
for both $\eta'$ and $\eta$ production.  We should note that 
the three-body $\eta^{(\prime)}$ yields are not very sensitive 
to $m_s$, unlike the two-body yields.    

In Table \ref{third} the two-body and three-body $CP$ 
averaged SM $Br$'s are summed for the range of parameters previously 
discussed.   
We find that ${\cal B}(B \to \eta' X_s)$ lies in the range
$(0.5 - 2.5) \times 10^{-4}$,
which is significantly smaller than the observed rate 
even for the largest $Br$'s.
The $CP$ asymmetries range from $2\%$ to $5\%$ with the larger asymmetries 
corresponding to the smaller $Br$'s.  In the case of $\eta$ 
production the SM $Br$ sums are slightly larger, lying in the range 
${\cal B}(B \to \eta X_s) \approx (0.7 - 2.6) \times 10^{-4}$,
with $CP$ asymmetries of $1.5 \% $ to $ 4.5 \% $. 
As already noted, we do not expect
significant changes in the ratio of $\eta$ to $\eta'$ 
yields when SM 2b-3b interference is taken into account.
Finally, with no momentum cut we obtain 
${\cal B}(B \to \eta' X_s ) \approx (0.25 - 1.0) 
\times 10^{-3}$ in the SM.  

\begin{figure}[t]
\centerline{
\epsfbox{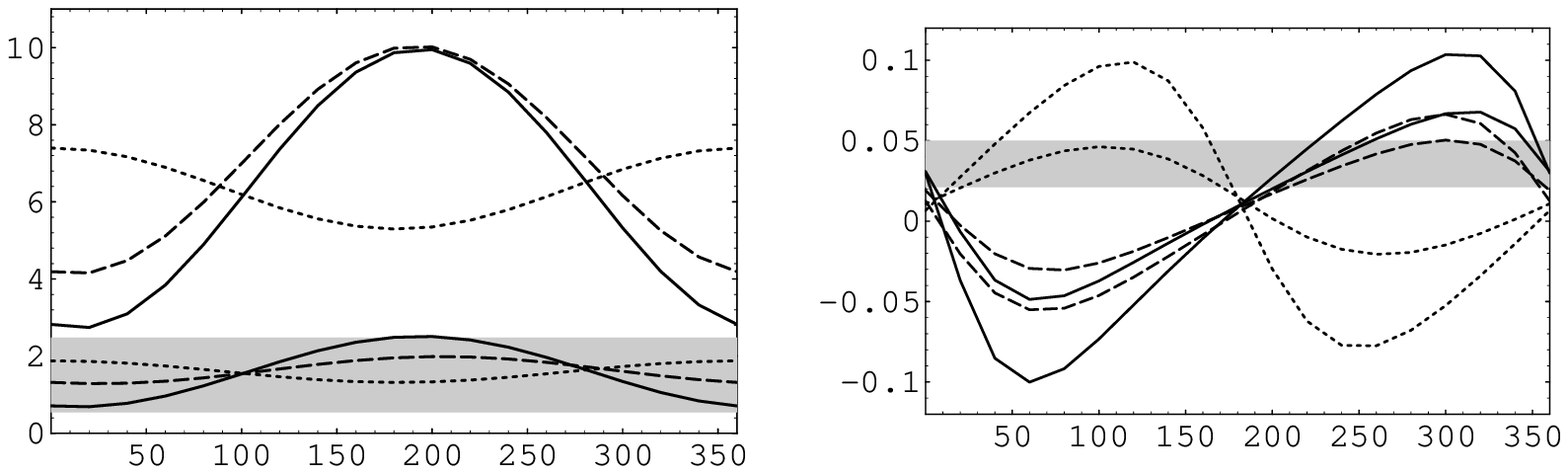}}
\caption {${\cal B}(B^- \to \eta' s \bar u)$ vs. $\theta$ and $CP$ 
asymmetries vs. $\theta$
for ${\cal B}(b \to sg) \approx 15\% $. Shaded regions are 
the corresponding SM ranges.  Solid curves are for $c_{11}' = 0$ 
($\theta = \theta_{11}$), 
long dashed curves are for  $|c_{11}| = |c_{11}'|$ and 
$\theta_{11}' = 180^o$ ($\theta = \theta_{11}$), short 
dashed curves are for $c_{11} = 0$ ($\theta = \theta_{11}'$).  
The larger $Br$'s (lower asymmetries) are for
$F_{1^-}^{\eta'} (0) = .5$, $f_8 = 1.25 f_{\pi}$, 
$\theta_{\eta \eta'} = -21^o$, $m_s = .1~GeV$, the lower $Br$'s 
(larger asymmetries) are for $F_{1^-}^{\eta'} (0) = .25$, $f_8 = f_{\pi}$, 
$\theta_{\eta \eta'} = -17^o$, $m_s = .2~GeV$. }
\label{five}
\end{figure}

Taking ${\cal B}(b \to sg) \approx 15\%$, 
the dependence of $\eta'$ yields 
on the relative magnitudes of $c_{11} $ and $c_{11}'$, corresponding phases,
and factorization model parameters is explored in Fig. 5.
Sums of pure two-body and pure three-body contributions have been plotted
for three illustrative cases:
$c_{11}' = 0$, $c_{11} = 0$, and $|c_{11}| = |c_{11}'|$.
It is worth noting that in 
models of quark mass generation a large hierarchy between $c_{11} $ 
and $c_{11}'$ is not expected in the absence of some special flavor symmetry.
According to Fig. (\ref{five}) factorization alone could 
well account for the observed 
inclusive yield in Eq. (\ref{cleoinc}) if dipoles were 
enhanced.\footnote{For $F_{1^-}^{\eta'} (0) = .5$ but $m_s = .15~GeV$,
$f_8 = f_{\pi}$, $\theta_{\eta \eta'} \approx - 17^o $
the maximum yields would be about 15\% lower than in Fig. (\ref{five}).} 
This conclusion persists when constraints from other rare decays are 
taken into account, as we'll see shortly.
The large range of $\eta'$ yields possible is in marked contrast to the 
low SM range, shown in grey.
The $\eta' $ $Br$ with no cut is $\sim (2.5 - 10) \times 10^{-3}$ for maximal 
constructive interference between dipole and penguin contributions.
The $\eta$ yields are about 10\% larger for 
$F_{1^-}^{\eta} (0) = F_{1^-}^{\eta'} (0) $
and  2b-3b interference not taken into account.

The $CP$ asymmetries for inclusive $\eta'$ production in Fig. (\ref{five})
are not much larger than the SM range from Table \ref{third}, despite 
the presence of significant new amplitudes with arbitrary weak phases. This 
is because the dominant three-body dipole contributions 
are weighted towards low $q^2$, where little or no
strong phase is generated for the interfering three-body penguin amplitudes.

\subsection{Exclusive $\eta^{(\prime)}$ production.}

It is straightforward to obtain factorization amplitudes for 
$B^\pm \to \eta^{(\prime)} K^{\pm} $ from $H_{eff}$ and $H_{11}$
in terms of the $\eta^{(\prime)}$ hadronic matrix elements 
listed in Eqs. (\ref{etapr:all}, \ref{eta:all}, \ref{betap:all}), 
the kaon matrix elements 
\begin{eqnarray}  
\langle K^- |\bar s \gamma^\mu \gamma_5 u |0 \rangle = - i 
\sqrt{2} F_K p_K^\mu \\
\langle K^-  |\bar s  \gamma_5 u |0 \rangle  =  - i 
\sqrt{2} F_K {m_{K^-}^2 \over {m_s + m_u} },
\end{eqnarray}  
and the analogs of the $B \to \eta'$ matrix elements 
in Eq. (\ref{betap:all}) for $B^- \to K^-$  transitions \cite{BSW}.
Again we consider the range $F_{1^-} (0) = .25 - .5$ for the 
$B \to \eta^{(\prime)}$ and $B \to K$ form factors at zero momentum transfer.  
Although we are looking at exclusive decays, individual quark momenta enter
the amplitudes because of $H_{11}$.  For simplicity we will 
identify $p_b$ with the $B$ meson momentum and make the usual quark 
model kinematical assumptions, taking
$p_s = p_{\bar s} = p_{\eta^{(\prime)}}/2 $ for non-spectator 
$\eta^{(\prime)}$ production
and $p_s = p_{\bar u} = p_{K^-} /2$ for spectator 
$\eta^{(\prime)}$ production. With these assumptions one obtains 
$q^2 = m_{B^-}^2  /2 - m_{\eta^{(\prime)}}^2/4 + m_{K^-}^2/2$
for the former and 
$q^2 = m_{B^-}^2  /2 - m_{K^-}^2/4 + m_{\eta^{(\prime)}}^2/2$ 
for the latter.

SM $Br$'s and CP asymmetries for $B^\pm \to \eta' K^\pm $
are summarized in Table \ref{fourth}. The $Br$'s range from 
$(1.1 - 12.1 )\times 10^{-5}$, 
with significant enhancement occurring at low values of $m_s$ 
due to the $1/m_s$ factors in $\langle \eta' |\bar s  \gamma_5 s |0 \rangle$  
and $\langle K^-  |\bar s  \gamma_5 u |0 \rangle$.
It is clear that SM factorization can reproduce the measured $Br$ in 
Eq. (\ref{cleoexc})
given a reasonably broad range of input parameters. 
The CP asymmetries can also be significant, varying from 4.5\% - 8 \%.
For $\eta$ production we find 
${\cal B}(B^\pm \to \eta K^\pm ) \approx (1.5 - 4.8 )\times 10^{-6}$ 
for $F_{1^-}^{\eta'} (0) = F_{1^-}^{\eta} (0) $,
in accord with the order of magnitude suppression expected relative to 
the $\eta' K^\pm$ yield from Lipkin's interference mechanism\cite{lipkin}.
The corresponding $CP$ asymmetries are large, at the 15\% - 20\% level.

\begin{figure}
\centerline{
\epsfbox{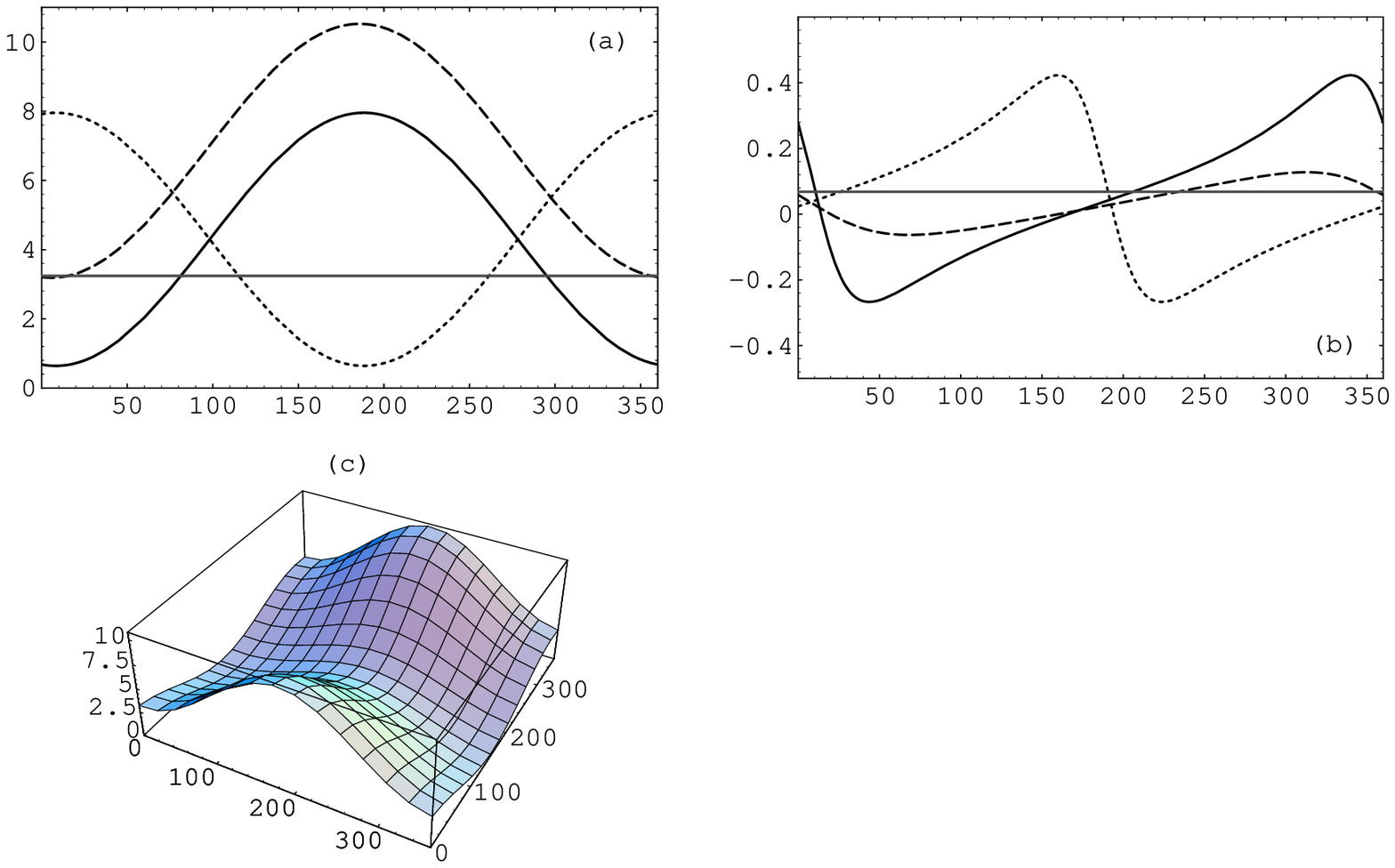}}
\caption {(a) $10^5 {\cal B}(B^\pm \to \eta' K^\pm)$ vs. $\theta$ 
and SM value (horizontal line) for 
$F_{1^-}^{\eta'}(0) = .35$, $F_{1^-}^K (0) = .38$, $m_s = .15~GeV$, $F_8 = F_\pi$,
$\theta_{\eta \eta '} = -17^o$.  Solid, long dashed, and short dashed curves are for 
same $c_{11}$ and $c_{11}'$ as in Fig. 5. 
(b) CP asymmetries vs. $\theta$ corresponding to (a). Horizontal line is SM value. 
(c) ${\cal B}(B^\pm \to \eta' K^\pm)$ in $(\theta_{11},\theta_{11}')$ 
plane for $|c_{11}| = |c_{11}'|$.}
\label{six}
\end{figure}

Enhanced dipole operators can lead to 
significant increases in the exclusive $\eta' $ yield, as is to be expected 
given the increases we saw are possible for the semi-inclusive yield.
Eq. (\ref{cleoexc}) can therefore be reproduced for a larger
region of factorization
model parameter space. This is illustrated in 
Fig. (\ref{six}), where exclusive $\eta'$ production 
with enhanced dipoles included is compared
to SM yields for intermediate choices of form factors and $m_s$. 
According to Fig. (\ref{six}) significantly larger $CP$ 
asymmetries than in the SM are also possible, although 
differences in magnitude  
decrease in regions of $\theta_{11}$ and/or $\theta_{11}'$ corresponding to 
large $\eta'$ production.
Finally, it is important to note that Lipkin's interference mechanism 
for exclusive decays persists in the case of enhanced $b \to sg$. 
For example, the $\eta$ yields corresponding to Fig. (\ref{six}) 
lie in the range 
${\cal B}(B \to \eta K^\pm) \sim (1.0 - 8.0)\times 10^{-6}$
if $F_{1^-}^{\eta'} (0) = F_{1^-}^{\eta} (0) $. 

\subsection{Comparison with other rare $B$ decays.}

It is important to check that large factorization model $\eta'$ yields 
obtained with enhanced $b \to sg$ are consistent with constraints 
from other rare decays.  
We have considered three processes: $B \to \phi X_s$, which is the most 
restrictive semi-inclusive decay, $B^\pm \to \phi K^\pm$, which is 
probably the most restrictive exclusive decay, and $B^\pm \to K^0 \pi^\pm$ 
which is not particularly restrictive but, like the other two, is 
least cumbersome to calculate because there are no tree-level
contributions (pure-penguin decays).  Recently the CLEO collaboration 
has presented the following bounds: 
${\cal B}(B \to \phi X_s) < 2.2 \times 10^{-4}$ (90\% ~c.l.) 
\cite{CLEOphiX}, 
${\cal B}(B^\pm \to \phi K^\pm ) <  1.2 \times 10^{-5}$ (90\%~ c.l.) 
\cite{CLEOphiK},
and ${\cal B}(B^\pm \to K^O \pi^\pm ) = 2.3^{+ 1.14}_{- 1.04} \times 10^{-5}$ 
\cite{CLEOhawaiialexander}.  
In the SM we obtain 
${\cal B}(B \to \phi X_s) \approx 1.1 \times 10^{-4}$ (also see 
\cite{deshheold,deshhe,ciuchini}), and ${\cal B}(B^\pm \to \phi K^\pm) 
\approx (0.32 - 1.3)  \times 10^{-5}$ as 
$F_{1^-}^{K} (0)$ is varied from .25 to .5.  
Neither $Br$ is sensitive to $m_s$.
For $F_{1^-}^\pi (0) = .333$ \cite{BSW} we obtain
${\cal B}(B^\pm \to K^O \pi^\pm ) = (1.0 - 2.1) \times 10^{-5}$ as 
$m_s$ is varied from .2 to .1 $GeV$. It is interesting that 
the largest  SM $\eta'$ yields in Tables \ref{third} and \ref{fourth} 
are consistent with the above constraints.
For the exclusive decays this is due to Lipkin's mechanism,  
whereas in the semi-inclusive case this is because there is no 
spectator contribution to $\phi$ production, whereas it constitutes the bulk
of the $\eta'$ yield. Another semi-inclusive process 
for which there is a spectator contribution, $B \to K_s X$,
is very weakly bounded, ${\cal B}(B \to K_s X) < 7.5 
\times 10^{-4}$ for $p_K > 2~GeV$ \cite{CLEOKep}.

\begin{figure}[t]
\centerline{
\epsfbox{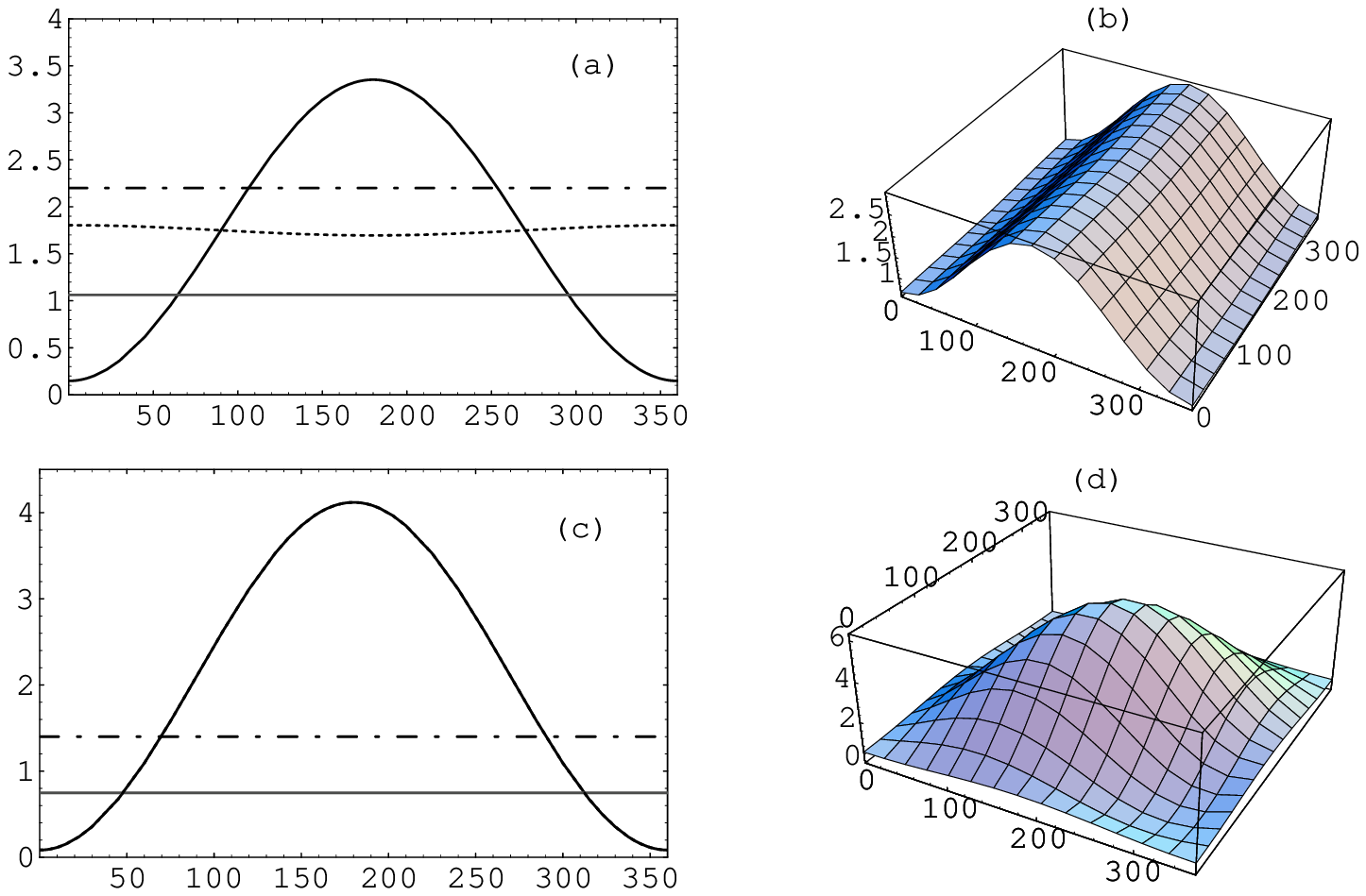}}
\caption {(a) $10^4 {\cal B}(B \to \phi X_s)$ vs. $\theta$ for ${\cal B}(b \to sg) \approx 15\%$.
Solid curve is for $c_{11}' = 0$ ($\theta = \theta_{11}$), dashed curve is for 
$c_{11} = 0$ ($\theta = \theta_{11}  '$), horizontal solid curve is SM value, and 
dot-dashed curve is the CLEO upper limit.  (b) ${\cal B}(B \to \phi X_s)$ 
in the $(\theta_{11},\theta_{11}  ')$ plane for $|c_{11}| = |c_{11} '|$.
(c) and (d) are analogous to (a) and (b) for $10^5 {\cal B}(B^\pm \to \phi K^\pm)$.  
In (c) solid and dashed curves coincide.}
\label{seven}
\end{figure}

In Fig. (\ref{seven}) we show the $\phi X_s$ and $\phi K^\pm$ yields with 
enhanced $b \to sg$ corresponding
to the examples in Fig. (\ref{six}).  Comparison of
Figs. (\ref{five} - \ref{seven}) shows that within the 
factorization model there is significant overlap of regions 
in ($c_{11}$, $c_{11}'$) space with 
exclusive and semi-inclusive $\eta' $ yields in the measured ranges 
and acceptable $\phi$ production.  
No additional constraints arise from $B^\pm \to K^0 \pi^\pm$.

\section{Exotic Mechanisms}

It is well known that the $\eta'$ is not a Goldstone boson -- its mass
does not vanish in the chiral limit $m_{u,d,s} \to 0$. This is believed to be
due to a large glue component in the $\eta'$. In the present context 
the enhanced glue content leads to a significant coupling
of the $\eta'$ to two gluons, which participates in $b \to s$ transitions.

It is important to notice that because of large cancellations \cite{wyler,liu},
the contribution of the $b \to s gg$ mode followed 
by a $gg \to \eta'$ transition
is extremely small. On the other hand, the ${\cal O}(1\%)$
$Br$ of the 
$b \to s g^*$ transition in the SM could, in principle,
be responsible for the anomalously large $b \to \eta' X_s$ decay rate.
This fact was exploited by AS \cite{atwood}, who considered the 
contribution from the $b \to s g^* \to \eta^\prime s g$ subprocess. They parameterized 
the effective $gg\eta'$ coupling as
\begin{equation}
V_{\mu \nu} \epsilon_1^\mu \epsilon_2^\nu =
H(q^2, k^2, q_{\eta^\prime}^2) \epsilon_{\alpha \beta \mu \nu}
q^\alpha k^\beta \epsilon_1^\mu \epsilon_2^\nu
\end{equation}
where $q^2 \equiv (p_b-p_s)^2$.
A constant form 
factor was assumed,
i.e.,
$H(q^2, k^2, q_{\eta'}^2) \simeq H(0, 0, m_{\eta'}^2) 
\equiv H_0$, and $H_0$ was extracted directly 
from the decay rate for $J/\psi \to \eta' \gamma$.
For
$\alpha_s(m_{J/\psi})=0.25$ and $\theta_{\eta \eta^\prime} = -17^o$,
the value $H_0 = 1.8 ~GeV^{-1}$ is obtained.  

With the central assumption of weak $q^2$ dependence in $H$, 
AS found that the observed $\eta^\prime$ yield could be fully accounted for.
However, it is clear that in order to obtain the total decay rate,
the differential distribution for this subprocess 
must be integrated over a wide range of
$q^2$, spanning approximately 1 $GeV^2$
to $m_b^2$. It is therefore of paramount importance to investigate 
the off-shell dependence of the form-factor describing the
$g^* \to \eta' g$ transition.

Determination of the exact $q^2$-dependence of 
$H(q^2,0, m_{\eta'}^2)$ is a difficult task.
We therefore employ a semi-phenomenological description of 
the form-factor, picking up the leading $q^2$-dependence and
parametrizing the rest by an appropriate constant.
To set the stage we recall the leading $q^2$-dependence of the
$\gamma^* \to \eta' \gamma$ form-factor in the limit of
large momentum transfer \cite{ter}. In this limit, 
a perturbative QCD description is possible and it is not difficult to
convince oneself that the simplest two-particle irreducible
kernel yields a form-factor that scales like $1/Q^2$ as 
$Q^2 \to \infty$ ($Q^2 = -q^2 $) \cite{ter}:
\begin{equation} \label{pqcd}
\Gamma_{\mu \nu} \epsilon_1^\mu \epsilon_2^\nu =
i e^2 F_{\eta'}(Q^2) \epsilon_{\alpha \beta \mu \nu}
q^\alpha k^\beta \epsilon_1^\mu \epsilon_2^\nu, ~~~
 F_{\eta'}(Q^2) = 4 \sqrt{\frac{2}{3}} \frac{F_\pi}{Q^2} \xi.
\end{equation}
with $\xi \simeq 0.5$. While the corresponding calculation for
$g^* \to \eta' g$ is more involved, the lowest order $Q^2$ 
dependence in the $Q^2 \to \infty$ limit should be the same,
so it is clear that the constant form factor 
approximation $H(q^2, 0, m_{\eta'}^2) \simeq H_0$ is 
not acceptable.

In order to obtain a more realistic parameterization we consider a 
model of the $g g \eta'$ vertex in which we couple a pseudoscalar 
current to two gluons through quark loops. This approximation
is believed to be a good one since the ``direct'' 
$gg \eta'$ coupling is suppressed -- the quantum 
numbers of $\eta'$ are $0^-$, while the lowest possible 
bound state of two gluons is $0^+$.
The calculation yields \cite{tri}
\begin{eqnarray} \label{vert}
V_{\mu \nu} &=& \sum_{f=u,d,s} \frac{a_f g_s^2 m_f}{2 \pi^2} 
\epsilon_{\alpha \beta \mu \nu} q^\alpha k^\beta
\int_0^1 dx \int_0^{1-x} 
\frac{dy}{m_f^2 - (1-x-y)(x q^2 + y k^2) - 
xy m_{\eta'}^2 - i\epsilon} 
\nonumber \\
&=& \sum_{f=u,d,s}\frac{4 \pi \alpha_s m_f}{2 \pi^2}              
\frac{a_f}{q^2 -m_{\eta'}^2} F^{(f)}_\triangle (q^2)
\epsilon_{\alpha \beta \mu \nu} q^\alpha k^\beta, ~~~~
F^{(f)}_\triangle (q^2) = I \Bigl ( \frac{m_{\eta'}^2}{m_f^2} \Bigr )-
I \Bigl ( \frac{q^2}{m_f^2} \Bigr ),
\end{eqnarray}
where the $I (x^2)$-functions are defined as
\begin{eqnarray}
I(x^2) = \left\{ \begin{array}{ll} 
-2 \mbox{arcsin} \frac{|x|}{2}, & 0 \leq x^2 \leq 4, \\
2 \Bigl[ \ln \Bigl (\frac{|x|}{2}+\sqrt{\frac{x^2}{4}-1}
\Bigr ) - \frac{i \pi}{2} \Bigr ]^2, & ~~~x^2 > 4.
\end{array}
\right. 
\end{eqnarray}
$a_f$ is the coupling of the $\eta'$ to 
the quarks in the loop, i.e., $a_f \eta^\prime \bar f \gamma_5 f$. One 
has to keep in mind that
this vertex is in general {\it non-local}, so that $a_f$ will 
depend on the momentum transferred through the vertex. Since this 
dependence is at most logarithmic, i.e. subleading, it
will be ignored in what follows.

Motivated by the above result
it is now reasonable to parametrize the effective $gg \eta'$
coupling as
\begin{equation} \label{vertex}
V_{\mu \nu} = - \frac{H_0 m_{\eta'}^2}{q^2 -m_{\eta'}^2}
\epsilon_{\alpha \beta \mu \nu} q^\alpha k^\beta, ~~~
\mbox{and}~~~
H (q^2,0,m_{\eta'}^2) = -\frac{H_0 m_{\eta'}^2}{q^2-m_{\eta'}^2}
\end{equation}
Note that $H_0$ {\it does} depend on $q^2$. 
We drop this dependence  
based on the following observations.
First of all, $q^2 > 4 m_f^2$ for any quark flavor over the whole range
of $q^2$ contributing to the total decay rate. This implies that 
$F^{(f)}_\triangle (q^2)$ depends on $q^2$ only logarithmically.
Also, as noted by HT \cite{hou2}, the factor 
$\alpha_s(\mu^2)$ implicit in $H_0$ must in general be running. 
They argued that the scale at which 
$\alpha_s$ is evaluated should be associated with the momentum transfered 
through the $gg \eta'$ vertex, introducing
another logarithmic dependence on $q^2$ in $H_0$. It is clear that these
dependences are subleading with respect to the strong power dependence
in Eq. (\ref{vertex}) and to first approximation can be
modeled by a constant $H_0$, which we identify with the value extracted from
$J/\psi \to \eta' \gamma$.

The strong $q^2$ dependence of Eq. (\ref{vertex}) implies additional 
suppression of the $b \to \eta' sg$ anomaly subprocess at large $q^2$. 
It is easy to estimate the resulting contribution to $b \to \eta' X_s$. 
The calculation is 
a straightforward generalization of
the one carried out by AS \cite{atwood} and HT \cite{hou2} so we do not include
any details here. We only 
point out that in the SM the interference between
the charge radius and dipole form factor contributions is {\it destructive}. 

Our SM result for the total branching ratio is
${\cal B}({b \to \eta' sg}) \approx 5.3 \times 10^{-5}$,
corresponding to ${\cal B}(b \to \eta' sg) 
\approx 1.6 \times 10^{-5}$ with experimental cut.  
The latter is more than an order of magnitude smaller than the observed 
$\eta^\prime$ yield in Eq. (\ref{cleoinc}). 
The recoil spectrum is shown in Fig. 8. 
As already pointed out in Refs. \cite{atwood,hou2},
the anomaly subprocess exhibits a rise with $m_{X_s}$ in the low $m_{X_s}$ 
region which is characteristic of three-body decays.  
The $q^2$ dependence of the $g g \eta'$ formfactor 
significantly distorts the $q^2$ distribution
so that the maximum is shifted towards lower values 
of $q^2$ compared to constant $g g \eta'$ formfactor.

In the case of enhanced $b \to sg$ we again consider
${\cal B}(b \to sg) \approx 15\%$, and for simplicity take $c_{11}^\prime = 0$.  
The resulting $Br$'s with experimental cut are presented 
in Fig. 9 as a funtion of $\theta_{11}^\prime$. 
These are more than an order of magnitude smaller
than obtained by HT under the assumption of mild $q^2$-dependence of the form factor. 
The corresponding CP violating asymmetries are also systematically 
smaller 
than those obtained by HT \cite{hou2}. This is because of the shift in 
the decay distribution towards lower values of $q^2$, leading to smaller strong phases 
in the charge radius form factor.

We also remark that the short-distance 
$b \to \eta' sg$ subprocess most probably {\it does not} affect 
the exclusive $B \to \eta' K$ branching ratios. 
If the gluon attaches itself to the $s$ quark then the exclusive decay
can be interpreted as being due to the short distance subprocess 
$b \to s gg$ which we already know is 
suppressed \cite{wyler}.  The other two possibilities, that the gluon attaches itself to 
the spectator quark, or that it contributes to the 
process via higher Fock states of the participating mesons,
must be additionally suppressed by the dynamics
of the process.

Another interesting explanation of the anomalously large
$\eta'$ production rate is to assume a large `intrinsic charm'
content for the $\eta'$ \cite{zhitnitsky}. 
Phenomenologically, this can be described as a mixing of 
the $\eta_c$ and $\eta'$. Since the value of the mixing angle is
obtained by a fit to the experimental branching ratios
for radiative charmonia decays to $\eta'$
this phenomenological estimate should already contain a contribution 
from the higher Fock states containing charm quarks \cite{chao}.
It is worth mentioning that intrinsic heavy quarks of the light 
mesons must be far from their mass shells in order to satisfy 
energy-momentum conservation. Therefore, NRQCD methods
developed for weakly bound quarks in charmonia can not be used to 
describe the dynamics of intrinsic
heavy quark states.

The heavy-light quark mixing for the $\eta-\eta'-\eta_c$ system
was considered in \cite{azimov} and found to be small, 
$\sin \alpha_P \simeq 2.4 \times 10^{-2}$. 
This leads to the estimate $Br (B \to \eta' X_s) \sim 
\sin^2 \alpha_P  Br (B \to \eta_c X_s) < 10^{-6}$ so that
the `intrinsic charm' content of the $\eta'$ is not
likely to be responsible for the observed large branching 
ratio. 
In addition to phenomenological estimates limiting the
size of the $\bar c c$ contribution, a large `intrinsic charm' explanation
for the $\eta^\prime$ yields
would lead to ${\cal B}(B \to \eta' K^* ) \sim 2 {\cal B}(B \to \eta' K ) $ 
and a quasi two-body momentum 
spectrum for $B \to \eta^\prime X_s$  \cite{zhitnitsky}, neither of which is observed. 

\section{Conclusions}

We have considered production of $\eta'$ in 
$B$ meson decays in the Standard Model and in models with enhanced $b \to sg$.
It is clear that SM factorization can, in principle, 
account for the exclusive $\eta' $ yield.  For a liberal range of
factorization model parameters the SM 
$B^\pm \to \eta' K^\pm$ branching ratio lies in the range
$(1.1 - 5.8) \times 10^{-5} $ for $m_s = .2~GeV$ and 
$(2.3 - 12.1)\times 10^{-5}$ for $m_s = .1~GeV$.  
Sizable $CP$ violating asymmetries at the 5\% - 8\% level are expected.

The inclusive yield requires a more complicated solution.
For the same range of parameters SM factorization  
gives ${\cal B}(B^\pm \to \eta' X_s  ) 
\sim (0.5 - 2.5) \times 10^{-4}$ including the experimental cut, 
with the largest yields corresponding to a fairly limited region 
of parameter space. In addition, the branching ratio of the QCD anomaly mediated subprocess
$b \to \eta^\prime s g$ considered in \cite{atwood}
is reduced to $\sim 1.6 \times 10^{-5}$ when 
taking proper account of the leading $q^2$-dependence 
of the $\eta^\prime gg$ coupling form factor, $H(q^2) \sim m_{\eta'}^2/(q^2-m_{\eta'}^2)$.
The large rate reported in \cite{atwood} for constant form factor
would require a factor of 50 enhancement due to non-perturbative effects, e.g., resonances, 
of the rate obtained with leading
$q^2$-dependence included - an unlikely possibility.
However,
a SM `cocktail' solution for large ${\cal B}(B^\pm \to \eta' X_s  )$
involving contributions from several mechanisms, e.g., factorization pushed to the limit (the largest
single contribution), large non-factorizable contributions, charmonia decays \cite{atwood}, 
`intrinsic charm' \cite{zhitnitsky}, and $b \to \eta^\prime s g $, is still a possibility.    

On the other hand, the intervention of Non-SM physics in the form of enhanced
chromomagnetic dipole operators provides a simple and elegant solution 
to the puzzle of large $\eta'$ production in $B$ decays. 
This explanation can be realized in a broad region of the available parameter space.
In fact, factorization model contributions
alone can account for both the semi-inclusive and exclusive $\eta^\prime$ yields without violating contraints from
other rare decays. The $b \to \eta^\prime s g$ rate, although significantly smaller, can 
be an order of magnitude larger than in the SM.  
$CP$ violating asymmetries in exclusive and semi-inclusive $B \to \eta^\prime$ decays 
due to new weak phases in the dipole operator coefficients 
can be larger than in the SM and of either sign.  However, large asymmetries in the 
`pure-penguin' decays $B^\pm \to \phi K^\pm, K^0 \pi^\pm$
would provide a more definitive signal for new physics because of the small asymmetries 
expected for these decays in the SM.

{\bf Acknowledgements.}  We would like to thank David Atwood, Tom Browder, Gerhard Buchalla,
John Donoghue, Gene Golowich, Barry Holstein, George Hou, Uli Nierste, Joao Soares
and Amarjit Soni for useful discussions.

\begin{table}
\begin{center}
\caption{Two-body contributions to
${\cal B}(B^\pm \to \eta' X_s )$ (average of CP conjugate decays) and 
corresponding $CP$ asymmetries
for (i) standard model without $O_{11}$, (ii) standard model with
$O_{11}$. $F_0 = F_{\pi}$. \label{first}}
\vspace*{0.5cm} 
\begin{tabular} {| c | c | c | } \hline
Inputs  & no $O_{11}$ & with $O_{11}$ \\ \hline
$m_s = .1$, $F_8 = F_\pi$, $\theta_{\eta \eta'} = -17^o $ & 
$ 3.7 \times 10^{-5}$, .51\%    & $3.2 \times 10^{-5}$, .59\%  \\
$m_s = .2$, $F_8 = F_\pi$, $\theta_{\eta \eta'} = -17^o $   
& $1.3 \times 10^{-5} $, 1.5\% & $1.04 \times 10^{-5}$, 1.8\% \\
$m_s = .1$, $F_8 = 1.25 F_\pi$, $\theta_{\eta \eta'} = -21^o $ & 
$8.4 \times 10^{-5} $, .21\% & $ 7.4 \times 10^{-5}$, .24\%  \\
$m_s = .2$, $F_8 = 1.25 F_\pi$, $\theta_{\eta \eta'} = -21^o $ & 
$2.8 \times 10^{-5} $, .63\% &  $2.34 \times 10^{-5}$, .75\% \\
\hline
\end{tabular}
\end{center}
\end{table}
 
\begin{table}
\begin{center}
\caption{Three-body contributions to 
${\cal B}(B^- \to \eta' X_s )$, with experimental cut,
versus $q^2_{min}$ [$GeV^2$] for  
(i) standard model with $O_{11}$, (ii) ${\cal B}(b \to sg) \approx 15\%$
with $|c_{11} | = |c_{11}'|$, $\theta_{11} = 180^o$, $\theta_{11}' = 0$.
$F^{\eta'}_1 (0) = .5$, $m_s = .1~GeV$. \label{second}}
\vspace*{0.5cm} 
\begin{tabular} {| c | c | c |  } \hline
$q^2_{min} $ &  SM  & ${\cal B}(b \to sg) \approx 15\%$\\ 
\hline
.4 &  $1.71 \times 10^{-4}$ & $12.1 \times 10^{-4}$  \\
.6  & $1.68 \times 10^{-4}$ & $9.61 \times 10^{-4}$ \\
.8 &  $1.66 \times 10^{-4} $ & $8.38 \times 10^{-4}$ \\
1.0 &   $1.65 \times 10^{-4}$ & $7.61 \times 10^{-4}$ \\
1.4 &  $1.64 \times 10^{-4}$ & $6.64 \times 10^{-4}$ \\
2.0 &  $1.62 \times 10^{-4}$ & $5.77 \times 10^{-4}$ \\
 \hline
\end{tabular}
\end{center}
\end{table}

\begin{table}
\begin{center}
\caption{Sum of two-body and three-body contributions to
${\cal B}(B^\pm \to \eta' X_s )$ (average of CP conjugate decays) in SM
for (i) $F_{1^-}^{\eta'} (0) = .25$, (ii) $F_{1^-}^{\eta'} (0) = .5$. 
$F_0 = F_{\pi}$. \label{third}}
\vspace*{0.5cm} 
\begin{tabular} {| c | c | c | } \hline
Inputs  & $F_{1^-}^{\eta'} (0) = .25$  & $F_{1^-}^{\eta'} (0) = .50$ \\ \hline
$m_s = .1$, $F_8 = F_\pi$, $\theta_{\eta \eta'} = -17^o $ & 
$ 7.6 \times 10^{-5}$    & $2.1 \times 10^{-4}$   \\
$m_s = .2$, $F_8 = F_\pi$, $\theta_{\eta \eta'} = -17^o $   
& $5.4 \times 10^{-5} $ & $1.8 \times 10^{-4}$ \\
$m_s = .1$, $F_8 = 1.25 F_\pi$, $\theta_{\eta \eta'} = -21^o $ & 
$1.2 \times 10^{-4} $  & $ 2.5 \times 10^{-4}$  \\
$m_s = .2$, $F_8 = 1.25 F_\pi$, $\theta_{\eta \eta'} = -21^o $ & 
$6.7 \times 10^{-5} $ &  $2.0 \times 10^{-4}$ \\
 \hline
\end{tabular}
\end{center}
\end{table}

\begin{table}
\begin{center}
\caption{
${\cal B}(B^\pm \to \eta' K^\pm )$ (average of CP conjugate decays) and 
corresponding $CP$ asymmetries [\%] in SM. $F_0 = F_{\pi}$.
\label{fourth}}
\vspace*{0.5cm} 
\begin{tabular} {| c | c | c | } \hline
Inputs  & $m_s = .1~GeV$ & $m_s = .2~GeV$ \\ \hline
$F_8 = F_\pi$, $\theta_{\eta \eta'} = -17^o $, $F^{\eta'}_1 (0) =
F^{K}_1 (0) = .25 $ & 
$ 2.3 \times 10^{-5}$, 5.4\%    & $1.1 \times 10^{-5}$, 8.0\%  \\
$F_8 = F_\pi$, $\theta_{\eta \eta'} = -17^o $, $F^{\eta'}_1 (0) =
F^{K}_1 (0) = .5 $  
& $9.1 \times 10^{-5} $, 5.4\% & $4.6 \times 10^{-5}$, 8.0\% \\
$F_8 = 1.25 F_\pi$, $\theta_{\eta \eta'} = -21^o $, $F^{\eta'}_1 (0) =
F^{K}_1 (0) = .25 $ & 
$3.0 \times 10^{-5} $, 4.5\% & $ 1.5 \times 10^{-5}$, 6.9\%  \\
$F_8 = 1.25 F_\pi$, $\theta_{\eta \eta'} = -21^o $, $F^{\eta'}_1 (0) =
F^{K}_1 (0) = .5 $ & 
$1.2 \times 10^{-4} $, 4.5\% &  $5.8 \times 10^{-5}$, 6.9 \% \\
\hline
\end{tabular}
\end{center}
\end{table}

\newpage
\newpage
\newpage
\newpage
\begin{figure}
\centerline{
\epsfbox{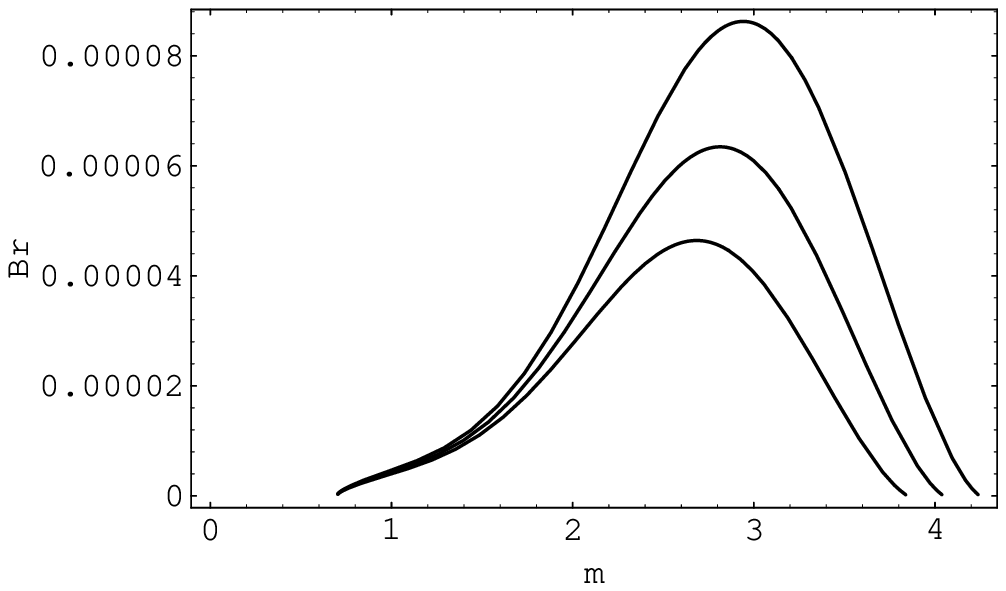}}
\caption {${{d {\cal B}}\over {d m_{X_s}} } (b \to \eta^\prime s g)$ vs. $m_{X_s}$
in the SM. Curves from top to bottom are for $m_b=5.2,~5.0$, and $4.8$ GeV.
$m_g = 0.5~GeV$ and $m_s = 0.2~GeV$.}
\label{spectrum}
\end{figure}
\newpage
\begin{figure}[bh]
\centerline{
\epsfbox{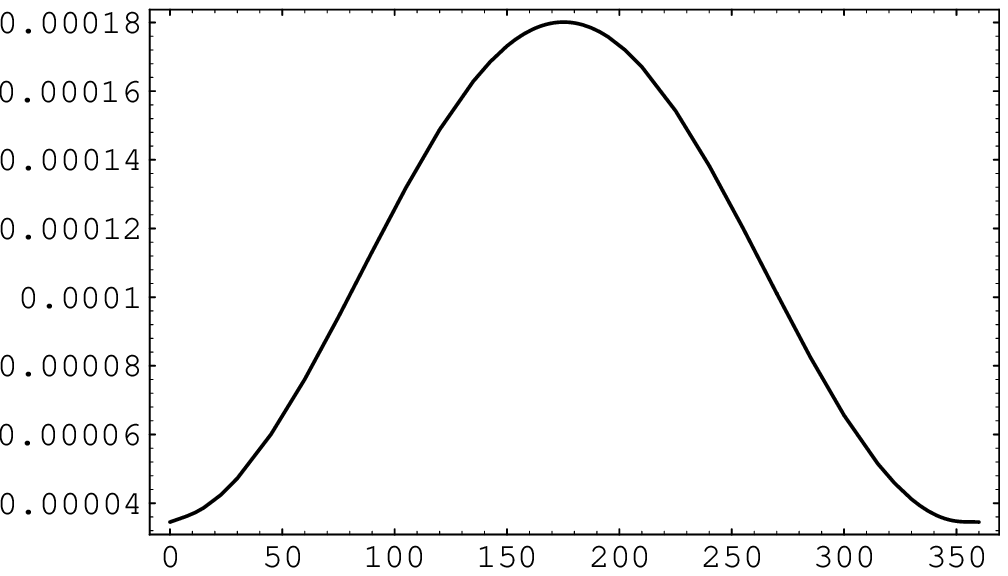}}
\caption {${\cal B}(b \to \eta' s g)$ vs. $\theta_{11}$
for ${\cal B}(b \to s g) \approx 15 \%$ and
$c_{11}'=0$.}
\label{bratio}
\end{figure}

\end{document}